\def\a{\alpha}\def\b{\beta}\def\c{\chi}\def\d{\delta}\def\e{\epsilon}
\def\f{\phi}
\def\l{\lambda}\def\m{\mu}\def\n{\nu}\def\o{\omega}\def
\p{\pi}\def\q{\psi}\def\r{\rho}\def\s{\sigma}
\def\y{\eta}\def\x{\xi}\def\z{\zeta}

\def\O{\Omega}\def\S{\Sigma}

\def\lie{{\cal L}}\def\de{\partial}\def\na{\nabla}
\def\id{\equiv}\def\ha{{1\over 2}}
\def\di{{\rm d}}

\def\st{spacetime }
\def\fe{field equations }
\def\tran{transformations }                                   
\def\bg{background }

\def\rep{representation }

\def\diff{diffeomorphisms }
\def\pb{Poisson brackets }\def\db{Dirac brackets }\def\ham{hamiltonian }
\def\cd{covariant derivative }\def\dof{degrees of freedom }
\def\hdim{higher dimensional }
\def\ads{anti-de Sitter }
\def\poi{Poincar\'e }
\def\des{de Sitter }
 
\def\GB{Gauss-Bonnet }\def\CS{Chern-Simons }
\def\EH{Einstein-Hilbert }

\def\section#1{\bigskip\noindent{\bf#1}\smallskip}
\def\nota{\footnote{$^\dagger$}}

\def\PL#1{Phys.\ Lett.\ {\bf#1}}\def\CMP#1{Commun.\ Math.\ Phys.\ {\bf#1}} 
\def\PRL#1{Phys.\ Rev.\ Lett.\ {\bf#1}} 
\def\PR#1{Phys.\ Rev.\ {\bf#1}}\def\CQG#1{Class.\ Quantum Grav.\ {\bf#1}} 
\def\NP#1{Nucl.\ Phys.\ {\bf#1}} 
\def\JMP#1{J.\ Math.\ Phys.\ {\bf#1}} 

 \def\IJMP#1{Int.\ J. Mod.\ Phys.\ {\bf #1}}

\def\ref#1{\medskip\everypar={\hangindent 2\parindent}#1}
\def\beginref{\begingroup
\bigskip
\centerline{\bf References}
\nobreak\noindent}
\def\endref{\par\endgroup}

\def\eA{\e_{ABCDE}\ }\def\emu{\e^{\m\n\r\s}\ }\def\ei{\e^{ijk}\ }
\def\Of{\O_{AB,CD}^{ij}}\def\Op{\O_{AB,C}^{i}}\def\Oab{\O_{\a\b}}
\def\AB{{AB}}\def\epsnum{\e_{a_1...a_{2n}}}                                             
\def\doff{degrees of freedom}


\magnification=1200
\baselineskip=18pt

{\nopagenumbers
\line{January 1998\hfil INFNCA-TH9801}
\vskip80pt
\centerline{\bf The dynamical structure of four-dimensional}
\centerline{\bf  Chamseddine's gauge theory of gravity}
\vskip40pt
\centerline{{\bf S. Mignemi}\nota{e-mail:MIGNEMI@CA.INFN.IT}}
\vskip10pt
\centerline {Dipartimento di Matematica, Universit\`a di Cagliari}
\centerline{viale Merello 92, 09123 Cagliari, Italy}
\centerline{and INFN, Sezione di Cagliari}
\vskip40pt
{\noindent 
We perform the Dirac hamiltonian analysis of a four-dimensional
gauge theory of gravity with an action of topological type, which
generalizes some well-known two-dimensional models. We show that,
in contrast with the two-dimensional case, the theory has a
non-vanishing number of dynamical degrees of freedom and that its structure
is very similar to higher-dimensional \CS gravity.}
\vskip60pt
P.A.C.S. Numbers: 04.20.Fy 04.60.-m 11.15.-q
\vfil\eject}

\section{1. Introduction}
The problem of finding a consistent quantum theory of gravity is one of the
most fascinating challenges in today's theoretical physics. Among the various
approaches
to the problem, many attempts have been done to write down general relativity
as a gauge theory, in the hope that what has been learned in the quantization
of gauge theories can be exploited also in the case of gravitation.
Some partial results have been obtained since the sixties in the direction
of identifying GR with some kind of gauge theory of the \poi group [1].
However it
has not been possible till now to do this in a completely satisfying way.
A closer equivalence of gravity with gauge theories can however be obtained in
lower dimensions [2-3]. It has been shown, in fact, that the 2- and
3-dimensional
analogue of GR can be written down as vector gauge theories of the
lower-dimensional (anti)-\des or \poi group with action of topological kind,
where by topological we mean that no background metric is introduced in the
formalism.

More precisely, in three dimensions one adopts a \CS action for the
(anti)-\des or \poi group [2]. After a suitable
identification of the components of the gauge connection with the dreibeins
and the Lorentz connection of the manifold, one recovers the three-dimensional
\EH action. An analogous mechanism works in two dimensions if the action is
chosen to be of the BF type [3].

These lower-dimensional models have in common the absence of local degrees
of freedom and this property renders their study much easier than
higher-dimensional
theories. In particular, one can perform their exact quantization in a
straightforward way [2-3]. It is therefore natural to ask whether
these models can be generalized to higher dimensions and if in this case
they acquire dynamical \dof and if their quantization can still be easily
achieved. Generalizations to higher dimensions are indeed possible, as was
first shown in [4], but the resulting theories can no longer be identified
with higher-dimensional general relativity.

The case of odd dimensions has been discussed in several papers [4-5]. 
The action is in this case the straightforward generalization to higher
dimensions of the \CS action,
and after the identification of the gauge fields with the geometrical
quantities gives rise to a gravitational action which is a sum with given
coefficients of \GB terms. (These generalized gravitational actions in higher
dimensions were first introduced by Lovelock [6]).

The even-dimensional case is less trivial. One possibility is to consider
\hdim BF theories: in this case the gauge fields must be coupled to
higher rank tensor fields [7].
A second possibility was suggested  by Chamseddine [4] and is perhaps closer
in spirit to the two-dimensional model.
According to his proposal, one proceeds as in two dimensions, and couples a
scalar multiplet to the field strength of the relevant gauge group.
After the usual identifications of gauge potentials and geometric quantities,
one still
obtains an action which is a sum of Euler densities, which however are now
coupled to the scalar fields. In addition, the action includes some further
terms involving products of
curvature and torsion. Some of the physical implications of these models have
been discussed in [8].

We again remark that in general these \hdim models are quite different
from GR, especially in the \poi case, where only the highest order \GB term
survives and hence no term proportional to the \EH action is included.

We have recalled that in three dimensions the \CS action for gravity possesses
no \doff. This is not true however for its \hdim generalizations, as was shown
in [9].
Analogously, the two-dimensional Chamseddine's lagrangian for gravitation does not
have dynamical \doff.
It would be interesting to know if this property extends to higher
dimensions. Investigations based on a perturbative expansion have shown that
this may depend on the specific \bg chosen for the calculation [8]. A deeper
understanding of the phase space of the model should however be obtained by
using the \ham formalism. This is the purpose of the present paper.

As usual with generally covariant
theories, we obtain a constrained hamiltonian system. The action is first-order
in the time derivatives and the hamiltonian results to be a linear combination
of the constraints. Adopting
the Dirac procedure [10], we separate the constraints in first class and second
class. The first class constraints generate the gauge \tran and spatial \diff.
Using standard methods [11], one can then calculate the number of dynamical
degrees of freedom. Our analysis is simplified by the fact that the \ham
formulation of our model displays many
similarities with that of \CS theories, which was discussed some
time ago in [9].

We shall consider only the case of $D=4$, but the discussion can be easily
extended to any even dimension. Also the generalization to more general gauge
groups may be obtained in a straightforward way.
We find that, like in the \CS theories, in higher dimensions the number of \dof
does not vanish.
A further analogy with the \CS theories is that the generator of the
time-like \diff is not independent from the other constraints.
This is a good feature in view of the quantization of the model, since it is
well
known that usually this is the constraint that is most difficult to solve.
On the other hand, the Dirac bracket structure is very involved in our case,
and we were not able to compute it explicitly.

\section{2. Gauge theories of gravity in 2{\it n} dimensions}
In a even number of dimensions, it is not possible to define a \CS action.
However, one can construct a different sort of action, which still
does not depend on any \bg metric and may therefore be called "topological".

In a $2n$-dimensional \st the most natural choices for the gauge group
of gravitation are
the \poi group $ISO(1,2n-1)$, the \des group $SO(1,2n)$ or the \ads group
$SO(2,2n-1)$, depending on the value of the cosmological constant $\l$,
which takes the value $\l=0$, $\l=1$ or $\l=-1$ respectively. 
The last two groups admit as invariant tensor the totally antisymmetric 
$(2n+1)$-tensor $\e_{A_1\dots A_{2n+1}}$, but the $\l=0$ case can be
easily recovered by In\"onu-Wigner contraction.

The generators $M_\AB$ of the gauge algebra satisfy the commutation relations
$$[M_{AB},M_{CD}]=\ha(h_{AD}M_{BC}-h_{AC}M_{BD}-h_{BD}M_{AC}+h_{BC}M_{AD})
\eqno(2.1)$$
with $h_{AB}=$ diag $(-1,1,\dots,1,\l)$ and the group indices $A, B,\dots$
run from 0 to $2n$.

As in standard Yang-Mills theory, local invariance under the gauge group
can be enforced by introducing a gauge connection one-form $A^{AB}$
with field strength 2-form $F^{AB}\id(dA+\ha[A,A])^{AB}=dA^{AB}+A^{AC}A^{CB}$,
where the indices are summed by means of $h_\AB$.
A gauge-invariant action of topological form can then be constructed by
taking the $2n$-form given by the exterior product of $n$ field strengths.
However, in order to construct a group invariant, one is forced to introduce a 
scalar multiplet $\y^A$ in the fundamental \rep of the group\nota{A different 
sort of 
generalization with an action linear in the $F^\AB$ can be obtained by 
introducing higher-rank forms [7].}.
The action can then be written as [4]
$$I=\int_{M_{2n}}\e_{A_1...A_{2n+1}}\ \y^{A_1}F^{A_2A_3}\dots F^{A_{2n}
A_{2n+1}}\eqno(2.2)$$
The field equations, obtained by varying the action with respect to the
scalar and to the gauge potential are given respectively by:
$$\eqalign{&\e_{A_1...A_{2n+1}}\ F^{A_2A_3}\dots F^{A_{2n}A_{2n+1}}=0\cr
&\e_{A_1...A_{2n+1}}\ F^{A_2A_3}\dots F^{A_{2n-2}A_{2n-1}}D\y^{A_1}=0\cr}
\eqno(2.3)$$
where $D$ is the gauge covariant derivative.

In order to establish a relation with $2n$-dimensional gravity,
one can now make the identifications
$A^{ab}=\o^{ab}$, $A^{a,2n}=e^a$, where $\o^{ab}=\o^{ab}_{\ \ \m}\di x^\m$ and 
$e^a=e^a_{\,\m}\di x^\m$ are the spin connection and vielbein 1-forms of
the $2n$-dimensional manifold and the indices $a,b...=0,\dots,2n-1$ refer to
the Lorentz subgroup $SO(1,2n-1)$ of the gauge group.
It follows that $F^{ab}=R^{ab}+\l\ e^ae^b$, $F^{a,2n}=T^a$, where 
$R^{ab}$ and $T^a$ are the curvature and the torsion 2-forms of the
$2n$-dimensional manifold, which are defined respectively as
$$R^{ab}=d\o^{ab}+\o^{ac}\o^{cb}\qquad\qquad T^a=de^a+\o^{ab}e^b\eqno(2.4)$$
and satisfy the Bianchi identities 
$$\eqalign{\na T^a&\id dT^a+\o^{ab}T^b=R^{ab}e^b\cr
\na R^{ab}&\id(dR+\o R-R\o)^{ab}=0}\eqno(2.5)$$
$\na$ being the Lorentz covariant derivative on the spacetime.
The scalar field $\y^A$ is also decomposed in a Lorentz scalar $\y\id\y^{2n}$ 
and a Lorentz vector $\y^a$.

With these identifications, the action (2.2) becomes
$$\eqalign{I=\int_{M_{2n}}\Big[&\epsnum\y(R^{a_1a_2}+\l\ e^{a_{1}}
e^{a_{2}})\dots (R^{a_{2n-1}a_{2n}}+\l\ e^{a_{2n-1}}e^{a_{2n}})\cr
+&2n\ \epsnum\y^{a_1}T^{a_2}(R^{a_3a_4}+\l\ e^{a_{3}}e^{a_{4}})\dots
(R^{a_{2n-1}a_{2n}}+\l\ e^{a_{2n-1}}e^{a_{2n}})\Big]\cr}
\eqno(2.6)$$
The first term in the action (2.6) represents a sum of \GB forms
multiplied by the scalar field $\y$ [5].
In the $\l=0$ case only the term $\y\,\e_{a_1...a_{2n}} R^{a_1a_2}\dots 
R^{a_{2n-1}a_{2n}}$ in the integral survives. This is the Euler class of the
manifold and
in the absence of the scalar field, it would be a total derivative.

In geometric form, the \fe (2.3) can be written as
$$\eqalign{\epsnum&(R^{a_1a_2}+\l\ e^{a_{1}}e^{a_{2}})\dots (R^{a_{2n-1}a_{2n}}
+\l\ e^{a_{2n-1}}e^{a_{2n}})=0\cr
\epsnum&T^{a_2}(R^{a_3a_4}+\l\ e^{a_{3}}e^{a_{4}})\dots (R^{a_{2n-1}a_{2n}}
+\l\ e^{a_{2n-1}}e^{a_{2n}})=0\cr
\epsnum&(R^{a_2a_3}+\l\ e^{a_2}e^{a_3})\dots (R^{a_{2n-2}a_{2n-1}}+\l\
e^{a_{2n-2}}e^{a_{2n-1}})(\na\y^{a_1}+\l\ \y\, e^{a_1})=0\cr
\epsnum&[2T^{a_2}(R^{a_3a_4}+\l\ e^{a_{3}}e^{a_{4}})\dots(R^{a_{2n-3}a_{2n-2}}
+\l\ e^{a_{2n-3}}e^{a_{2n-2}})(\na\y^{a_1}+\l\ \y\, e^{a_1})\cr
&\qquad+(R^{a_1a_2}+\l\ e^{a_{1}}e^{a_{2}})\dots (R^{a_{2n-3}a_{2n-2}}+\l\
e^{a_{2n-3}}e^{a_{2n-2}})(\na\y-\y^be_b)]=0\cr}$$

\section{3. The 4-dimensional theory}
In the following, we shall concentrate our attention on four dimensions.
The gauge group is given in this case by $ISO(1,3)$, $SO(1,4)$ or $SO(2,3)$
with generators $M_{AB}$, $A,B=0,\dots,4$ and commutation relations (2.1),
where now $h_\AB$ = diag($-1,1,1,1,\l$).

The components of the gauge field strength are given by
$$F^{AB}_{\m\n}=\de_\m A^{AB}_\n-\de_\n A^{AB}_\m+A^{AC}_\m A^{CB}_\n-
A^{AC}_\n A^{CB}_\m\eqno(3.1)$$
where $A^{AB}_\m$ ($\mu=0,\dots,3$) is the gauge connection.

The four-dimensional Chamseddine action can be written as
$$I=\int_{M_4}d^4x\ \eA\emu\y^A F^{BC}_{\m\n}F^{DE}_{\r\s}\eqno(3.2)$$
This action is invariant, up to boundary terms, under the standard gauge
\tran with parameter $\c^\AB$:
$$\d_G A^{AB}_\m=-D_\m\c^{AB}\qquad\qquad\d_G\y^A=\c^{AB}\y^B\eqno(3.3)$$
where $D_\m$ is the gauge \cd. 
The action is also invariant under \diff of the \st manifold $M$ with
parameter $\x^\m$:
$$\d_D A^{AB}_\m=\lie_\x A^{AB}_\m=\x^\n\de_\n A^{AB}_\m+A^{AB}_\n\de_\m\x^\n
\qquad\qquad\d_D\y^A=\lie_\x\y^A=\x^\m\de_\m\y^A\eqno(3.4)$$
where $\lie_\x$ is the Lie derivative in the direction of $\x^\m$.
It is useful to define improved \diff [12], that differ from (3.4) by a gauge 
transformation with parameter $\c^\AB=\x^\n A^{AB}_\n$. One has
$$\d_I A^{AB}_\m=\x^\n F^{AB}_{\n\m}
\qquad\qquad\d_I\y^A=\x^\m\de_\m\y^A+\x^\m A^{AB}_\m\y^B=\x^\m D_\m\y^A
\eqno(3.5)$$
Varying the action (3.2) with respect to $\y^A$ and $A^{AB}_\m$ one obtains
the \fe
$$\eqalign{&\eA\emu F^{BC}_{\m\n} F^{DE}_{\r\s}=0\cr&\eA\emu
F^{BC}_{\m\n}D_\r\y^A=0\cr}\eqno(3.6)$$
In the following we shall make repeated use of the Bianchi identities
$$D_{[\n}F_{\r\s]}=\emu D_\n F_{\r\s}=0\eqno(3.7)$$

In order to perform the Hamiltonian analysis, we assume that the \st
manifold has topology $R\times\S$ and decompose the 1-form $A^{AB}$ as
$$A^{AB}_\m dx^\m=A^{AB}_0 dt+A^{AB}_i dx^i\qquad (i=1,2,3)$$
The action can then be decomposed as
$$I=\int_R\int_\S dtd^3x\left[ l_{AB}^i(\y,A)\dot A^{AB}_i+A^{AB}_0 K_{AB}
(\y,A)\right]\eqno(3.8)$$
with
$$\eqalignno{&l_{AB}^i=\eA\ei F^{CD}_{jk}\y^E&(3.9)\cr
&K_{AB}=\eA\ei F^{CD}_{jk}D_i\y^E=D_i l_{AB}^i&(3.10)\cr}$$
The \fe (3.6) are decomposed accordingly as follows:
$$\eqalignno{&\eA\ei(\dot A^{BC}_i -D_i A^{BC}_0)F^{DE}_{jk}=0&(3.11)\cr
&\eA\ei(2\dot A^{CD}_j D_k\y^E+F^{CD}_{jk}\dot\y^E-2D_jA^{CD}_0D_k\y^E+
F^{CD}_{jk}A^{EF}_0\y^F)=0&(3.12)\cr &K_{AB}=0&(3.13)\cr}$$
which follow from the variation of $\y^A$, $A^{AB}_i$, $A^{AB}_0$
respectively. The last equation can be interpreted as a constraint, with the
non-dynamical fields $A^{AB}_0$ playing the role of Lagrange multipliers.

\section{4. Analysis of the constraints}
The action (3.8) is first order in the time derivatives. The \ham analysis for
first-order actions can be performed either using the formalism introduced by
Fadeev and Jackiw [13], which postulates an unusual type of brackets in
configuration space, or by means of the standard Dirac method for constrained
systems [10]. Due to the presence of second class constraints, it is more
convenient in our case to make recourse to the Dirac formalism (see
ref. [14] for a general discussion).
Therefore, since the action is linear in the time derivatives of $A^{AB}_i$
and does not contain the time derivatives of the $\y^A$, in addition to the
$K_\AB$, we must impose the 35 constraints
$$\eqalign{\f_{AB}^i&\id p_{AB}^i-l_{AB}^i\approx0\cr
\q_A&\id\p_A\approx0\cr}\eqno(4.1)$$
where $\f_{AB}^i$ and $\p_A$ are the momenta canonically conjugate to
$A^{AB}_i$ and $\y^A$ respectively.
The \pb between these constraints are
$$\eqalign{\Of &\id\{\f_{AB}^i,\f_{CD}^j\}={\d l_{CD}^j\over\d A^{AB}_i}
-{\d l_{AB}^i\over\d A^{CD}_j}=-2\eA\ei D_k\y^E\cr
\Op &\id\{\f_{AB}^i,\q_C\}=-{\d l_{AB}^i\over\d\y^C}=\eA\ei F^{DE}_{jk}\cr
\O_\AB &\id\{\q_A,\q_B\}=0\cr}\eqno(4.2)$$

It is also convenient to replace the constraints $K_{AB}$ with new constraints
$G_{AB}$, which generate the gauge \tran (3.3):
$$-G_{AB}=K_{AB}+D_i\f_{AB}^i+\y_{[A}\q_{B]}=D_ip_{AB}^i+\y_{[A}\p_{B]}$$
Indeed, it is easy to check that
$$\eqalign{&\d A^{AB}_i=\left\{A^{AB}_i,\int_\S\c^{CD}G^{CD}\right\}=
-D_i\c^{AB}\cr
&\d\y^A=\left\{\y^A,\int_\S\c^{BC}G^{BC}\right\}=\c^{AB}\y^B\cr}\eqno(4.3)$$
One has
$$\eqalign{&\{G_{AB},G_{CD}\}=\ha(h_{AD}G_{BC}-h_{AC}G_{BD}-h_{BD}G_{AC}+
h_{BC}G_{AD})\cr
&\{\f^i_{AB},G_{CD}\}=\ha(h_{AD}\f^i_{BC}-h_{AC}\f^i_{BD}-h_{BD}\f^i_{AC}
+h_{BC}\f^i_{AD})\cr
&\{\q_A,G_{BC}\}=\ha(h_{AB}\q_C-h_{AC}\q_B)\cr}\eqno(4.4)$$
It follows that the $G_\AB$ are first class and their \pb reproduce the gauge
algebra (2.1).

The Hamiltonian density is now
$${\cal H}=A^{AB}_0G_{AB}+u^{AB}_i\f_{AB}^i+v^A\q_A\eqno(4.5)$$
where $u^{AB}_i$ and $v^A$ are Lagrange multipliers enforcing the constraints 
$\f_{AB}^i$ and $\q_A$.
The consistency condition $\dot G_{AB}\approx0$ is automatically satisfied
because the $G_{AB}$ are first class, while the other consistency conditions
$$\eqalign{\dot\f_{AB}^i=& \{\f^i_{AB},H\}\approx u^{CD}_j\Of+v^C\Op=0\cr
\dot\q_C=& \{\q_C,H\}\approx u^{AB}_i\Op=0\cr}\eqno(4.6)$$
give restrictions on the Lagrange multipliers $u^{AB}_i$, $v^C$. Hence, no
further constraints appear.

We have already seen that the constraints $G_\AB$ are first class.
In order to investigate the nature of the constraints $\f_\AB^i$, $\q_B$, 
one must consider the matrix $\Oab$ formed with the \pb of the constraints 
[14],
where $\a$, $\b$ stay for the indices $a$, $b$, $i$, etc.
$$\Oab=\pmatrix{\{\f_{AB}^i,\f_{CD}^j\}&\{\f_{AB}^i,\q_F\}\cr
\{\q_E,\f_{CD}^j\}&\{\q_E,\q_F\}\cr}\eqno(4.7)$$

It turns out that this matrix is not invertible on the constraint surface
and therefore some combinations of the constraints $\f_\AB^i$, $\q_A$ are
first class. To show this, let us find the null eigenvectors $V_\b$ of 
$\Oab$, using the relations (4.2). One must solve the matrix equation
$$\Oab V_\b=\pmatrix{\Of&\O_{AB,F}^i\cr -\O_{CD,E}^j&0\cr}\pmatrix{V^{CD}
_j\cr V^F\cr}=0\eqno(4.8)$$
This yields
$$\eqalignno{&-\O_{CD,E}^jV^{CD}_{(l)j}=\eA\ei F^{AB}_{jk}V^{CD}_{(l)i}=0
&(4.9)\cr
&\Of V^{CD}_{(l)j}+\O_{AB,F}^iV^F_{(l)}=-\eA\ei[2V^{CD}_{(l)j}D_k\y^E+
F^{CD}_{jk}V_{(l)}^E]=0&(4.10)\cr}$$
The first equation admits the three solutions $V^{CD}_{(l)i}=F^{CD}_{li}$,
with $l=1,2,3$. Substituting in the second, one gets $V^E_{(l)}=D_l\y^E$.
This is indeed a consequence of the identity
$$\eA\ei[2F^{CD}_{lj}D_k\y^E+F^{CD}_{jk}D_l\y^E]=\d^i_lK_\AB\approx0
\eqno(4.11)$$

Hence, the $35\times 35$ matrix $\Oab$ has at least three null eigenvectors
$$V_{(l)\b}=\pmatrix{F^{CD}_{li}\cr D_l\y^E\cr}\eqno(4.12)$$
which correspond to first class constraints. These are given explicitly by
$$H_l=\f_{AB}^i F^{AB}_{li}+\q_AD_l\y^A=p_{AB}^i F^{AB}_{li}+\p_AD_l\y^A$$
and generate the improved spatial \diff (3.5). In fact,
$$\d A^{AB}_i=\left\{A^{AB}_i,\int_\S H_l\x^l\right\}=\x^lF^{AB}_{li}
\qquad\qquad\d\y^A=\left\{\y^A,\int_\S H_l\x^l\right\}=\x^lD_l\y^A
\eqno(4.13)$$

It is interesting to note that, in contrast with the two-dimensional case [2],
these constraints are in general independent of the constraints $G_\AB$,
generating local gauge transformations. A similar situation arises in
\CS theories, where the dependence occurs only in three dimensions, but not
in higher dimensions [9].

Another important analogy with \CS theories is that the generator
of time \diff is a linear combination of the first class constraints $G_\AB$
and $H_l$, since
this symmetry is not independent from the other ones. This can be proved by
showing that on-shell the time \diff can be written as space \diff with
suitable parameters. The proof goes essentially like in the \CS case [9]:
indeed, the \fe (3.11), (3.12) can be written in terms of the matrix $\Oab$
as
$$\Oab F_{0\b}=0\eqno(4.14)$$
where
$$F_{0\b}=\pmatrix{F^{CD}_{0i}\cr D_0\y^E\cr}\eqno(4.15)$$
Hence, if $\Oab$ has only the three null eigenvectors $V_{(l)\b}$, then
some parameters $\z^l$ must exist such that $F_{0\b}=\z^l V_{(l)\b}$.
Thus, for a time diffeomorphism, parametrized by $\x^\m=(\x^0,0)$, (3.5) can
be written as
$$\d_I\pmatrix{A^{CD}_i\cr \y^E\cr}=
\x^0\pmatrix{F^{CD}_{0i}\cr D_0\y^E\cr}=
\x^0\z^l\pmatrix{F^{CD}_{li}\cr D_l\y^E\cr}\eqno(4.16)$$
which is a space diffeomorphism with parameter $\x^0\z^l$.
Analogously, if further null eigenvectors
are present, the time \diff can be written as a linear combination
containing also the generators of the corresponding symmetries.

\section{5. Degrees of freedom count}
We are finally in a position to compute the number of local \dof of the
theory.
If the only null eigenvectors of $\Oab$ are the three vectors $V_{\b(l)}$
obtained above,
the count goes as follows: one has 70 canonical variables ($A^{AB}_i$,
$\y^A$, $p_{AB}^i$, $\p_A$), 10 first class constraints $G_\AB$ associated
with gauge invariance,
3 first class constraints $H_i$ associated with spatial diffeomorphism 
invariance and $35-3=32$ second class constraints.
The number ${\cal N}$ of local degrees of freedom is therefore given by
(see e.g [11])
$${\cal N}= \ha(P-2F-S)=6$$
where $P$ is the dimension of phase space and $F$ and $S$ are the number of
first and second class constraints, respectively.

Of course this is the maximum possible number of \doff, which is reached if
the matrix $\Oab$ has no further null eigenvectors besides the $V_{\b(l)}$ 
and these are linearly independent.
The validity of these conditions depends on the
region of the phase space one is considering. For example, in the region
corresponding to maximally symmetric \st ($F^\AB=0$) and vanishing $\y^A$, 
the matrix $\Oab$
becomes null and hence all constraints are first class and no \dof are left.
This is in accordance with the results found in [8] by a perturbative
expansion.

In order to prove the existence of regions of the phase space where the
number of \dof is maximal, one should check if there are explicit solutions
of the constraints such that the conditions above are satisfied.
Although, due to the complexity of the matrix $\Oab$, we were not able to
check this for explicit solutions, we find plausible that the conditions of
maximality hold for generic solutions, since no accidental symmetries, which
would give rise to further gauge invariances and hence to more null
eigenvectors, seem to be present in the action.

To complete the \ham analysis one should still compute the \db, which
permit one to get rid of the second class constraints. For two phase space
functions $A$ and $B$, these are given in general by
$$\{A,B\}^*=\{A,B\}-\int_\S dz\{A,\f_\a(z)\}J_{\a\b}(z)\{\f_\b(z),B\}$$
where  $\f_\a$ are the second class constraints and $J_{\a\b}$ is the inverse
of the matrix $\bar\O_{\a\b}$ formed with the \pb $\{\f_\a, \f_\b\}$.
In general, for first order systems,
one obtains non-trivial brackets between the fields [13,14]. We expect a similar
situation to arise also in our case. However, we are not able to compute
explicitly the \db, since we lack an explicit expression for extracting the 32
independent second class constraints out of (4.1).

Once one has obtained the \db, one can proceed to the quantization of the
theory. However, due to the non-trivial structure of the constraint algebra,
it still seems difficult to obtain a Hilbert space realization for it.

\section{6. Conclusions}
We have studied the \ham dynamics of a gauge model with an action of
topological form, which can be identified with a theory of gravity
in four dimensions and generalizes some well-known two-dimensional models.
We have shown that this model displays many similarities with the
odd-dimensional \CS theories.
The action is first order in the time derivatives and the \ham analysis can
be performed using the Dirac analysis of constrained systems.
The theory admits first class constraints related to gauge and \diff
invariance, and a set of second class constraints. In particular, the
generator of time \diff is not independent from the other constraints.
The computation of the local \dof shows that in contrast with the
two-dimensional case, their number does not vanish.
Unfortunately, it is not easy to explicitly separate the first class from
the second class constraints and then to calculate the \db. This is quite
disappointing in view of a possible quantization of the model.

It would be interesting to classify the local \dof in terms of their spin.
This can be more easily achieved in a perturbative approach. Preliminary
results indicate that in the riemannian limit (vanishing torsion), one has
a spin-2 excitation (graviton) in the (anti)-\des case, and two scalars in
the \poi case. The remaining \dof of the full theory are of course due to the
torsion.

Finally, we note that our investigations could easily be extended to higher
dimensions and to different gauge groups, not directly related to gravitation.

\section{Acknowledgements}
{\noindent I wish to thank L. Garay for a valuable discussion.}

\beginref
\ref [1] R. Utiyama, \PR{101}, 1597 (1956);
T.W.B. Kibble, \JMP{2}, 212 (1961);
\ref [2] T. Fukuyama and K. Kamimura, \PL{B160}, 259 (1985);
K. Isler and C. Trugenberger, \PRL{63}, 834 (1989);
A.H. Chamseddine and D. Wyler, \PL{B228}, 75 (1989);
\ref [3] A. Ach\'ucarro and P.K. Townsend, \PL{B180}, 89 (1986);
E. Witten, \NP{B311}, 46 (1988); \NP{B332}, 113 (1989);
\ref [4] A.H. Chamseddine, \PL{B233}, 291 (1989); \NP{B346}, 213 (1990);
\ref [5] F. M\"uller-Hoissen, \NP{B346}, 235 (1990);
M. Ba\~nados, C. Teitelboim and J. Zanelli, \PR{D49}, 975 (1994);
\ref [6] D. Lovelock, \JMP{12}, 498 (1971);
\ref [7] G.T. Horowitz, \CMP{125}, 417 (1989);
\ref [8] S. Mignemi, \CQG{14}, 2157 (1997); \CQG{15}, 289 (1998);
\ref [9] M. Ba\~nados, L.J. Garay and M. Henneaux, \PR{D53}, R593 (1996);
\NP{B476}, 611 (1996);
\ref [10] P.A.M. Dirac, {\it Lectures on quantum mechanics}, Yeshiva Un. Press,
New York 1964;
\ref [11] M. Henneaux, C. Teitelboim and J. Zanelli, \NP{B332}, 169 (1990);
\ref [12] R. Jackiw, \PRL{41}, 1635 (1978);
\ref [13] L. Fadeev and R. Jackiw, \PRL{60}, 1692 (1988);
\ref [14] J. Govaerts, \IJMP{A5}, 3625 (1990).

\endref
\end